\documentclass[11pt, twocolumn]{article}

\usepackage{amsfonts}
\usepackage{amsmath}
\usepackage{amssymb}
\usepackage{mathrsfs}

\usepackage{graphicx}
\usepackage{afterpage}
\usepackage{verbatim}
\usepackage{afterpage}
\usepackage{fancybox}
\usepackage{color}

\usepackage{amsthm}
\usepackage[normalem]{ulem}
\usepackage{bm}

\usepackage{enumitem}

\usepackage[procnames]{listings}
\usepackage{color}
\usepackage{textcomp}

\definecolor{keywords}{RGB}{255,0,90}
\definecolor{comments}{RGB}{0,0,113}
\definecolor{red}{RGB}{160,0,0}
\definecolor{green}{RGB}{0,150,0}
\definecolor{gray}{gray}{0.5}

\usepackage{tikz}
\usetikzlibrary{spy,calc}

\usetikzlibrary{shapes.geometric, arrows}
\usetikzlibrary{arrows, decorations.markings}

\newtheorem{theo}{Theorem}[section]

\newcommand{\bb}[1]{{\bm #1}}

\newcommand{\Radon}{\mathscr{R}}
\newcommand{\Back}{\mathscr{B}}

\newcommand{\R}{\mathbb{R}}

\usepackage{authblk}
\title{\bf Low-complexity Distributed Tomographic Backprojection for large datasets}
\author[1]{Gilberto Martinez Jr.\thanks{FAPESP 2016/16238-4}}
\author[1]{Janito V. Ferreira Filho}
\author[1]{Eduardo X. Miqueles\thanks{CNPQ 442000/2014-6}}
\affil[1]{Brazilian Synchrotron Light Laboratory (\textsc{lnls})/\textsc{CNPEM}, Campinas, SP, Brazil}
\date{}                     
\begin{document}


\maketitle

\begin{abstract}
In this manuscript we present a fast \textsc{gpu} implementation for tomographic reconstruction of large datasets using data obtained at the Brazilian synchrotron light source. The algorithm is distributed in a cluster with 4 \textsc{gpu}'s through a fast pipeline implemented in \textsc{c} programming language.  Our algorithm is theoretically based on a recently discovered low complexity formula, computing the total volume within $O(N^3\log N)$ floating point operations; much less than traditional algorithms that operates within $O(N^4)$ \textit{flops} over an input data of size $O(N^3)$. The results obtained with real data indicate that a reconstruction can be achieved within 1 second provided the data
is transferred completely to the memory.
\end{abstract}


\section{Introduction}

\begin{figure}[h]
\centering
\includegraphics[scale=0.5]{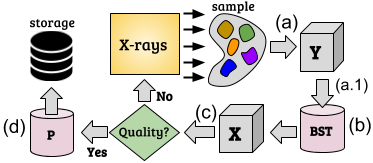}
\caption{Fluxogram for a fast tomographic scan.}
\label{fig:flowSirius}
\end{figure}

In this manuscript, we present a fast implementation of the well-known \emph{filtered-backprojection algorithm} (\textsc{fbp}) \cite{kak_slaney,natterer,deans}, which has the ability to reproduce reliable image reconstructions in a reasonable amount of time, before taking further decisions. The \textsc{fbp} is easy to implement and can be used to take fast decisions about the quality of the measurement, i.e., sample environment, beam-line conditions, among others. Figure \ref{fig:flowSirius} shows the fluxogram for an ideal tomographic experiment.
A synchrotron facility able to measure a three-dimensional dataset $\bm Y$ within few seconds, needs a fast reconstruction algorithm able to provide a fast "preview" of the tomography within the same amount of time (Fig.\ref{fig:flowSirius}.(a.1)). If the experimental conditions are not satisfactory, the quality of the reconstruction will decrease, and the researcher can decide either to make another scan, or to process later the data using advanced reconstruction algorithms or even high quality segmentation methods (Fig.\ref{fig:flowSirius}.(d)). The difficulty here is that the \textsc{fbp} algorithm consist basically in two mathematical operators, which are filtering and backprojection. Filtering is \textsc{fft} based since is a low-pass convolution operation \cite{deans}. Backprojection, on the other hand, is defined as an average through all the x-rays passing at a given pixel; therefore presenting high computational complexity of $O(N^3)$ for an image of $N^2$ pixels. The brute-force approach to compute the backprojection operator can be made extremely slow, even using a \textsc{gpu} implementation. Sophisticated ray-tracing strategies can also be used to make the running time faster and others analytical strategies reduce
the backprojection complexity to $O(N^2\log N)$, see \cite{andersson,bresler2}.
Our approach to compute the backprojection is called \textsc{bst} - as an acronym
to \emph{backprojection-slice theorem} \cite{miqueles} - having the same low complexity 
of $O(N^2 \log N)$ although easier to implement than his competitors, producing less
numerical artifacts and following a more traditional "gridding strategy" similar as \cite{marone}. There are
several others reconstruction softwares reproducing quasi-real time reconstructions, \cite{gursoy2014tomopy, marone2017towards, shkarin2015gpu, mirone2010pyhst}. 

The computational gain of \textsc{bst} over the brute-force approach for computing the backprojection
operator is presented using a fast pipeline for the data access. Our results indicate 
that reconstructions through \textsc{bst} can be performed within 1 second using the \textsc{ibm}/\textsc{minsky} (4 \textsc{nvidia} P100) for datasets of size $2048\times 2048 \times 2048$. A comparison is made with a 
\textsc{sgi/tesla} using 4 \textsc{nvidia} K80. To make a reliable comparison, all codes were implemented without taking advantages of the \textsc{nvidia}-\textsc{nvlink}. Also, we have tested our algorithm at small dimensions using only one \textsc{gpu}. Devices like \textsc{jetson} TX1, indicates that a fast reconstruction is possible for images of size $512\times 512 \times 512$, a conventional \textsc{gpu} like \textsc{gt}740M usually coupled with domestic notebooks can handle datasets of size $1024\times 1024 \times 1024$ and finally a \textsc{titan}-X coupled
with a standard \textsc{pc} handle volumes of size $2048\times 2048\times 2048$. In this sense, the reconstruction package supply three different aspects of a tomographic experiment: i) live reconstruction at the beamline assisting fast decisions by the researcher, ii) advanced reconstructions after the measurement using the beamline \textsc{gpu} power iii) reconstructions of the dataset without the beamline \textsc{gpu} power, at conventional desktops/notebooks. 

\begin{figure}[b]
\centering
\begin{tabular}{ccc}
    (a) & (b) \\
    \begin{tikzpicture}    
    \def\zlength{-0.5cm}
    
    \foreach \zangle [count=\i from 0] in {70}{
    \begin{scope}[shift={({mod(\i,2)*4cm},{-floor(\i/2)*4cm})}, 
        x=(0:1cm), y=(90:1cm),z=(\zangle:\zlength)]
    
        \def\sliceZ{0.8}
        \def\side{2}
    
        \filldraw[color=gray!40] (0,\sliceZ,0) -- (0,\sliceZ,\side) -- 
        (\side,\sliceZ,\side) -- (\side,\sliceZ,0) -- cycle;
        \draw[dashed] (0,\sliceZ,0) -- (0,\sliceZ,\side) -- 
        (\side,\sliceZ,\side) 
        -- (\side,\sliceZ,0) -- cycle;
    
        \draw (\side,0,0) -- (\side,\side,0) node[midway,right] {$\{s_k\}$} -- 
        (0,\side,0);
        \draw (0,0,\side) -- (\side,0,\side) node[midway,below] {$\{t_i\}$} -- 
        (\side,\side,\side) -- (0,\side,\side) -- (0,0,\side);
        \draw (\side,0,0) -- (\side,0,\side) node[midway,below right] 
        {$\{\theta_j\}$};
        \draw (\side,\side,0) -- (\side,\side,\side);
        \draw (0,\side,0) -- (0,\side,\side);
    
        \node[cm={1,0,cos(\zangle),sin(\zangle),(0,0)}] at (1,\sliceZ,1){$\bm y(t,\theta)$};
    \end{scope}
    }
    \end{tikzpicture}
    &
    \begin{tikzpicture}
    \def\zlength{-0.5cm}    
    \foreach \zangle [count=\i from 0] in {70}{
    \begin{scope}[shift={({mod(\i,2)*4cm},{-floor(\i/2)*4cm})}, 
        x=(0:1cm), y=(90:1cm),z=(\zangle:\zlength)]
    
        \def\sliceZ{0.8}
        \def\side{2}
    
        \filldraw[color=gray!40] (0,\sliceZ,0) -- (0,\sliceZ,\side) -- 
        (\side,\sliceZ,\side) -- (\side,\sliceZ,0) -- cycle;
        \draw[dashed] (0,\sliceZ,0) -- (0,\sliceZ,\side) -- 
        (\side,\sliceZ,\side) 
        -- (\side,\sliceZ,0) -- cycle;
    
        \draw (\side,0,0) -- (\side,\side,0) node[midway,right] {$\{s_k\}$} -- 
        (0,\side,0);
        \draw (0,0,\side) -- (\side,0,\side) node[midway,below] {$\{u_1\}$} -- 
        (\side,\side,\side) -- (0,\side,\side) -- (0,0,\side);
        \draw (\side,0,0) -- (\side,0,\side) node[midway,below right] 
        {$\{u_2\}$};
        \draw (\side,\side,0) -- (\side,\side,\side);
        \draw (0,\side,0) -- (0,\side,\side);
    
        \node[cm={1,0,cos(\zangle),sin(\zangle),(0,0)}] at (1,\sliceZ,1){$\bm x(u_1,u_2)$};
    \end{scope}
    }
    \end{tikzpicture}   
\end{tabular}
\caption{Definition of three-dimensional datasets: (a) Measured $\bf Y$ 
(b) Reconstructed $\bf X$.}
\label{fig:dataset}
\end{figure}
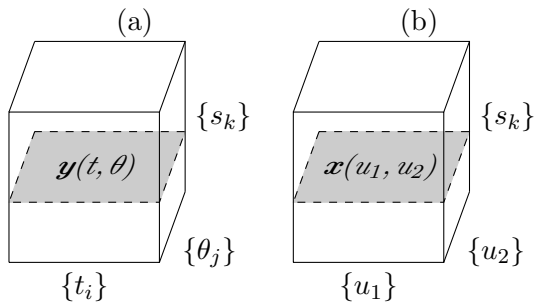

\begin{figure}[h]
\centering
\tikzstyle{application} = [rectangle, minimum width=0.5cm, rounded corners, minimum height=0.5cm, text centered, draw=black] 
\tikzstyle{arrow} = [thick,->]
\tikzstyle{arrow2} = [thick,<->]                                                                          \begin{tikzpicture}
\node (sino) [application] {$\bm y$\includegraphics[scale=0.2]{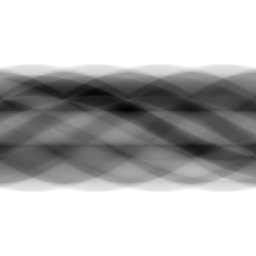}};
\node (feat) [application, above of=sino, yshift=1.5cm] {$\bm x$\includegraphics[scale=0.2]{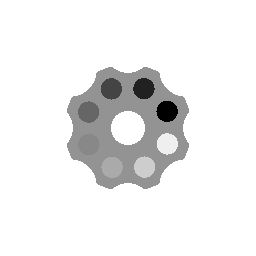}};
\node (back) [application, right of=sino, xshift=2.0cm] {$\bm b$\includegraphics[scale=0.2]{back_sino}};
\node (fsino) [application, right of=feat,xshift=2.0cm] 
{$\bm h$\includegraphics[scale=0.2]{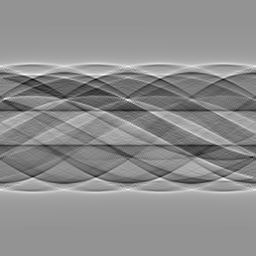}};
\draw [arrow] (sino) --  node [above] {$\mathcal B$} (back); 
\draw [arrow] (feat) -- node [left] {$\mathcal R$} (sino); 
\draw [arrow] (fsino) -- node [above] {$\mathcal B$} (feat);
\draw [arrow] (sino) -- node [above] {$\mathcal F$} (fsino);
\end{tikzpicture}
\caption{Action of operators Radon $\Radon$, Backprojection $\Back$ and Filtering $\mathcal F$.}
\label{fig:action}
\end{figure}

\medskip
\noindent{\bf Description of the problem}: We consider the transmission tomographic problem using x-rays generated at synchrotron facility based on parallel x-rays. The Radon transform has been used extensively as the mathematical object modelling the inverse problem. A typical tomographic measure provides
a three-dimensional dataset $\bf Y$ as presented in Figure \ref{fig:dataset}.a. Dataset $\bf Y$ is such that each plane $s \times t$ determines a radiography of the sample, for a constant projection angle $\theta_j$ varying on a discrete mesh with $V$ points from $0$ to $\pi$. The plane $s\times t$ is a discretization of a \textsc{ccd} camera with $N\times N$ pixels. The pair $(N,V)$ is a characteristic of the imaging beamline, typically $N \sim 2048$ and $V > 1001$, which means that $\bf Y$ is a large dataset with approximately $9$GiB. The tomographic problem for parallel rays is posed in the following manner: find a reconstructed three-dimensional dataset $\bf X$ from $\bf Y$ in such
a way that a given slice ($s_k$ constant) $\bm x(u)=\bm x(u_1,u_2)$ of the cube $\bf X$ 
is related to the same slice $\bm y(t,\theta)$ from volume $\bf Y$ through the linear operation
\begin{equation} \label{eq:radon}
\bm y(t,\theta) = \Radon \bm x(t,\theta) = \int_{\R^2} \bm x(u) \bm \delta(t- u\cdot \xi_\theta) \mathrm d u
\end{equation}
with $\xi = (\cos \theta,\sin\theta)$. Equation (\ref{eq:radon}) is the
Radon transform \cite{deans} from the two-dimensional function $\bm x$. 
Inverting the operator $\Radon$ is the mathematical core of most tomographic problems. There are several numerical algorithms for this task. One of the most celebrated algorithms is known as \emph{filtered-backprojection}, given by
the following formula $\bm x = \Back [\mathcal F \bm y]$ where $\Back$ is the 
\emph{backprojection operator} - the adjoint of 
$\Radon$ - defined as the following integral operator
\begin{equation} \label{eq:back}
\bm b(u) = \Back[\bm y](u) = \int_0^\pi \bm y(u\cdot \xi_\theta, \theta) \mathrm d\theta
\end{equation}
The operator $\mathcal F$ is a convolution, acting only on the
first variable $t$, i.e., $\bm h(t,\theta) = \mathcal F \bm y(t,\theta) = \bm y(t,\theta) \star \ell(t)$
with $\hat{\ell}(\sigma) = |\sigma|$. Figure \ref{fig:action} illustrate 
the filtered-backprojection action, together with $\Back$, $\Radon$ and 
$\mathcal F$ on a point source function $\bm x(u) = \bm \delta(u - a)$ (for a random point $a\in \R^2$). The function $\bm y= \Radon \bm x$ is often referred to as a sinogram because the Radon transform of an off-center point source is a sinusoid. The backprojection operation simply propagates the measured sinogram back into the image space along the projection paths.
Under physical assumptions that
are beyond the scope of this manuscript, the photon propagation
through a sample obeys the Lambert-Beer law $\bm I(\eta) = \bm I_0(\eta) e^{-\bm y(\eta) },\ \ \ \eta=(t,\theta)$ with $\bm I,\bm I_0$ standing for the transmitted and incident 
photon counting on the pixel camera, parameterized by the point 
$(t,\theta)$ and over the slice $s_k$. The filtered backprojection algorithm
applied over the sinogram $\bm y$ (\textsc{fbp}) offers a good reconstruction under strictly severe conditions on the measured data $\{\bm I,\bm I_0\}$, almost never satisfied at real measurements. There are three main processing steps on the data,
before reconstruction, they are:

\noindent{\bf(a)} {\it Normalization}: the sinogram $\bm y$ is obtained theoretically using the logarithmic function. In practice, a dark/flat field correction is used $\bm y = -\log \left[ (\bm I - \bm D)/(\bm I_0 - \bm D) \right]$, where $\bm D$ is the dark-field measurement (i.e., without the sample). 
\noindent{\bf (b)} {\it Centering}: In practice, the tomographic device suffers from several mechanical imprecisions, most of them, intrinsic to the problem. In this case, we obtain an experimental sinogram which is a not an exact realization of a Radon transform. This is case of a sample rotating above a precision stage, which is not perfectly aligned to the camera vertical axis. Here, the center of rotation of the sample is unknown, and the originated sinogram is slightly shifted from the correct center of rotation. As a consequence, the measured sinogram $\bm y$ is a noisy and shifted version of the theoretical one $\bm y(t - \beta, \theta)$
where $\beta$ is the unknown shift. 

\noindent{\bf (c)} {\it Ring Filtering}: Dead camera pixels or imperfections on the scintillator
remain constant as the angle $\theta$ varies within the interval $[0,\pi]$, giving origin to constant artifacts on the gathered frames $\bm I$. Therefore, strong stripes arise in
the sinogram $\bm y$, producing concentric rings on the final reconstructed image. There are also several ring artifacts correction algorithms, producing an  approximation of $\bm y$ as discussed in \cite{miquelesRings}.

\section{Fast Backprojection} 

The backprojection corresponds to an average 'smear' of all
projections passing through a single pixel $\bm x$. Since an average is described as 
a convolution, it is natural to regard formula (\ref{eq:back}) as an average in the frequency domain. It was shown recently \cite{miqueles} that the operator $\Back \bm y$ can be computed within $O(N^2 \log N)$ flops per slice, assuming that $\bm y$ is a sinogram having dimension of order $O(N^2)$. This is not the only fast approach to compute $\Back$, many others can be found in \cite{bresler2,andersson} while other algorithms as \cite{marone, potts} propose to reconstruct $\bm x$ using $\bm y$ as an input, through the well known \emph{Slice Theorem} or \emph{Fourier-Slice Theorem} (\textsc{fst}) 
relating the values of $\bm y$ and $\bm y$ in the reciprocal space. 
Our approach, the so-called \emph{Backprojection-slice Theorem} (\textsc{bst}) is 
based on the following theorem: 
\begin{theo}
	 Let $\bm y$ be a given sinogram and\/ $\widehat{\cdot}$ denotes the Fourier transform operation. The backprojection $\bm b = \Back \bm y$ can be computed through
	\begin{equation} \label{eq:bst}
	\widehat{\bm b}(\sigma\cos \theta, \sigma\sin\theta) = \hat{\bm y}(\sigma,\theta) / \sigma
	\end{equation}
	with $\sigma>0 \in \R$ and $\theta\in[0,2\pi]$.
\end{theo}

The \textsc{bst} approach is the dual of the \textsc{fst}, in the sense that 
$\bm y= \Radon \bm y$ and $\bm b = \Back \bm y$
are related in the reciprocal space within a polar grid. If $\bm y=\Radon \bm x$ the computation of $\bm b=\Radon \bm x$ follows easily convolving the feature image $\bm x$ with the point-spread function $1/\|u\|_2$ \cite{deans}, but $\bm x$ is unknown for pratical experiments. Also, \textsc{fst} provide a direct link between $\{\bm x,\bm y\}$ 
and $\{\bm b,\bm y\}$. If $\bm y\not=\Radon \bm x$ the link between $\{\bm x,\bm y\}$ is no longer valid through \textsc{fst} but we can still provide a connection between $\{\bm b, \bm y\}$. This is the main goal of the \textsc{bst} approach. The straight usage of the Fourier transform for the implementation of (\ref{eq:bst}) produce big artifacts near the origin, caused by the fact that the values on sinogram on the line $t=0$ are not equal to $0$. This problem can be solved with usage of short-time Fourier transform near the
origin, with window $w$, e.g. Kaiser-Bessel window function. 
The \textsc{bst} strategy applied over a sinogram image $\bm y$ is obtained after 7 processing stages $\{P_k\}$, i.e. $\bm b = P_6 P_5 P_4 P_3 P_2 P_1 P_0 (\bm y)$.
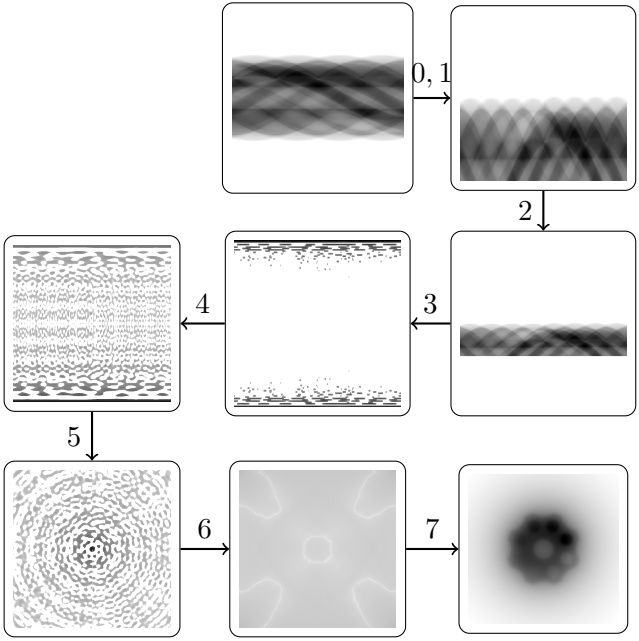
\begin{figure}[h]
\centering
\tikzstyle{backprojector} = [circle, dashed, minimum width=1cm, rounded corners, minimum height=1cm, text centered, draw=black] 
\tikzstyle{application} = [rectangle, minimum width=0.5cm, rounded corners, minimum height=0.5cm, text centered, draw=black] 
\tikzstyle{arrow} = [thick,->]
\tikzstyle{arrow2} = [thick,<->]
\begin{tikzpicture}
\node (kernel) [application] {\includegraphics[scale=0.08]{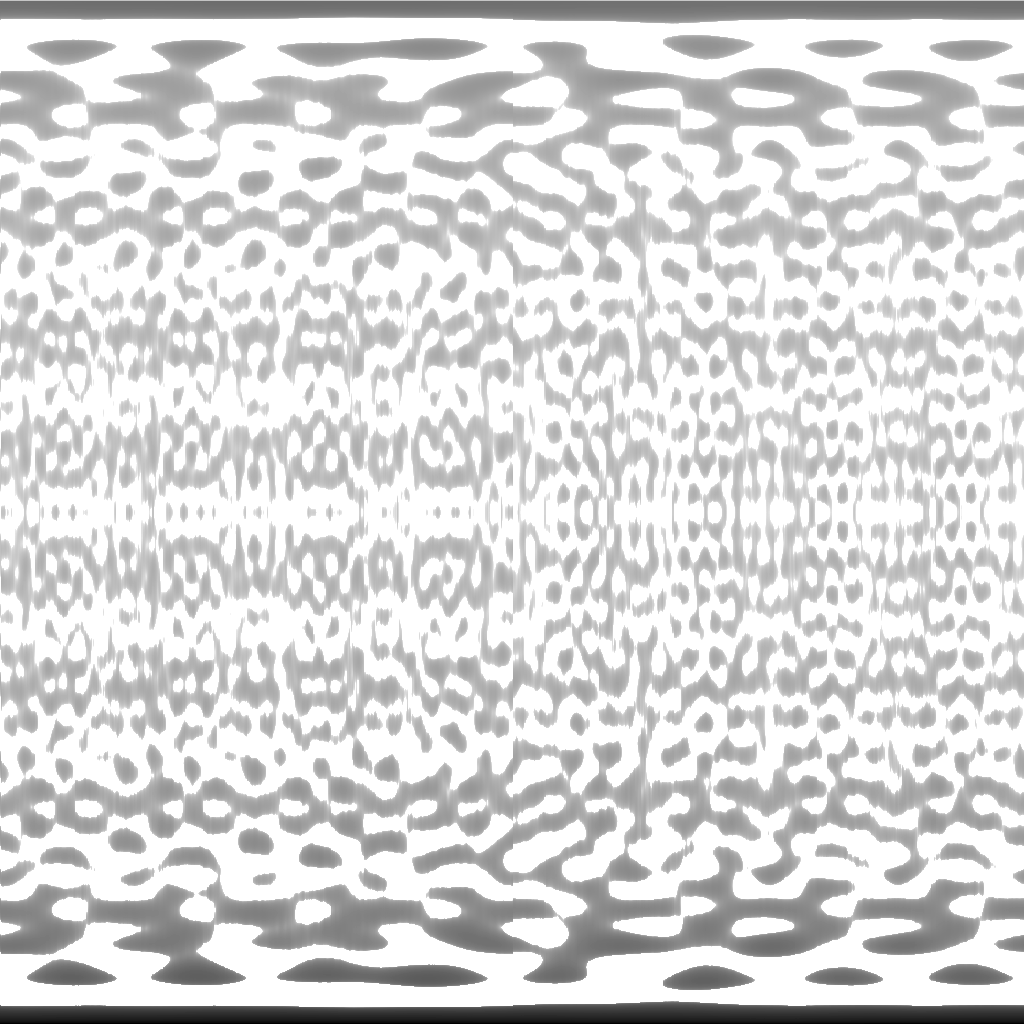}};

\node (1Dfft) [application, right of=kernel, xshift=2.0cm] {\includegraphics[width=2.2cm,height=2.2cm]{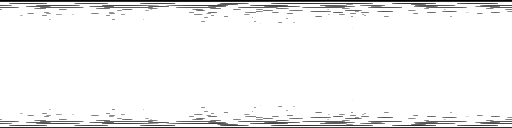}};

\node (bessel) [application, right of=1Dfft, xshift=2.0cm] {\includegraphics[width=2.2cm,height=2.2cm]{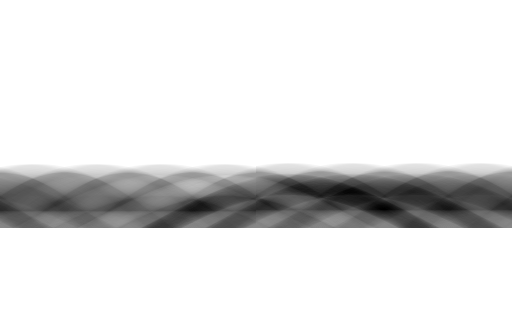}};


\node (Psino) [application, above of=1Dfft,yshift=2.0cm] 
{\includegraphics[scale=0.35]{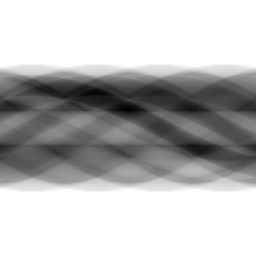}};

\node (zeroP) [application, right of=Psino, xshift=2.0cm] {\includegraphics[width=2.2cm,height=2.2cm]{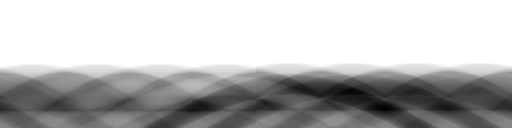}};

\node (interp) [application, below of=kernel, yshift=-2.0cm] {\includegraphics[scale=0.08]{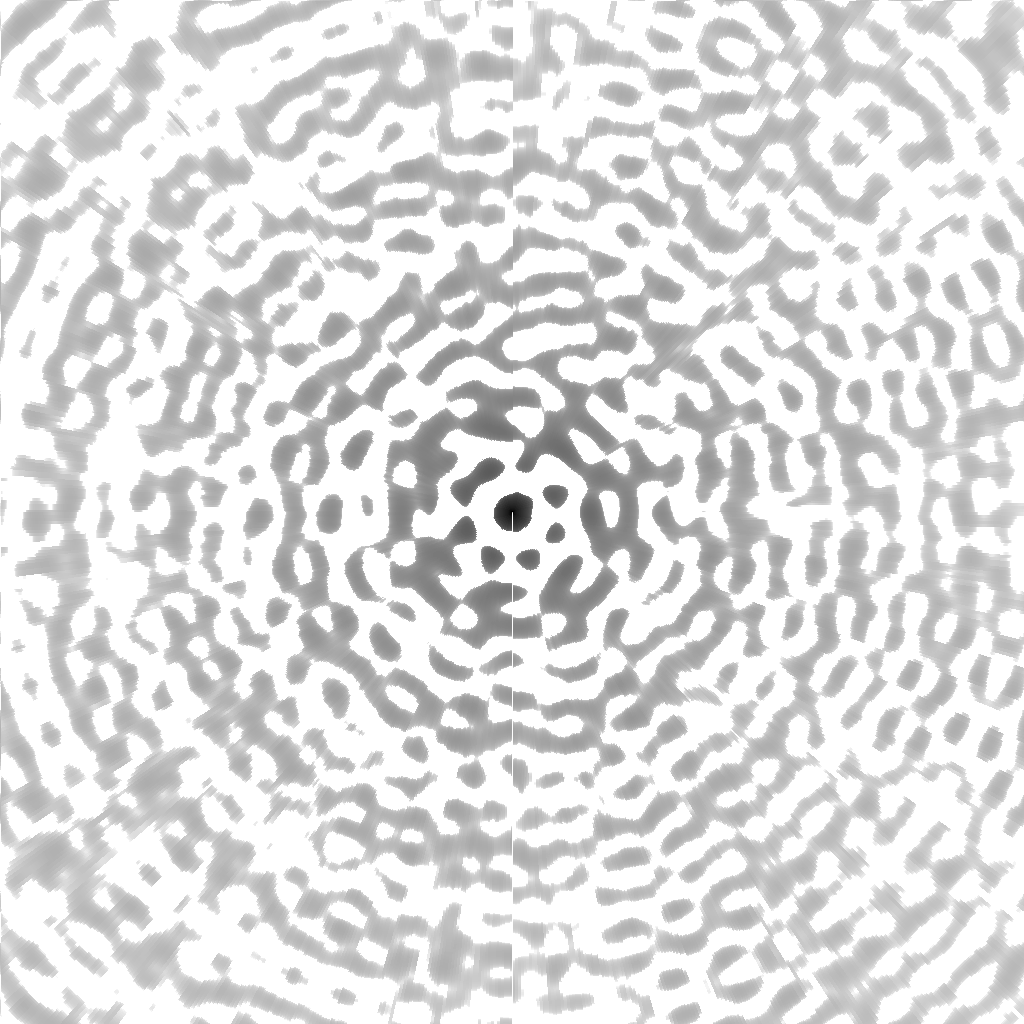}};

\node (2Dfft) [application, right of=interp, xshift=2.0cm] {\includegraphics[scale=0.08]{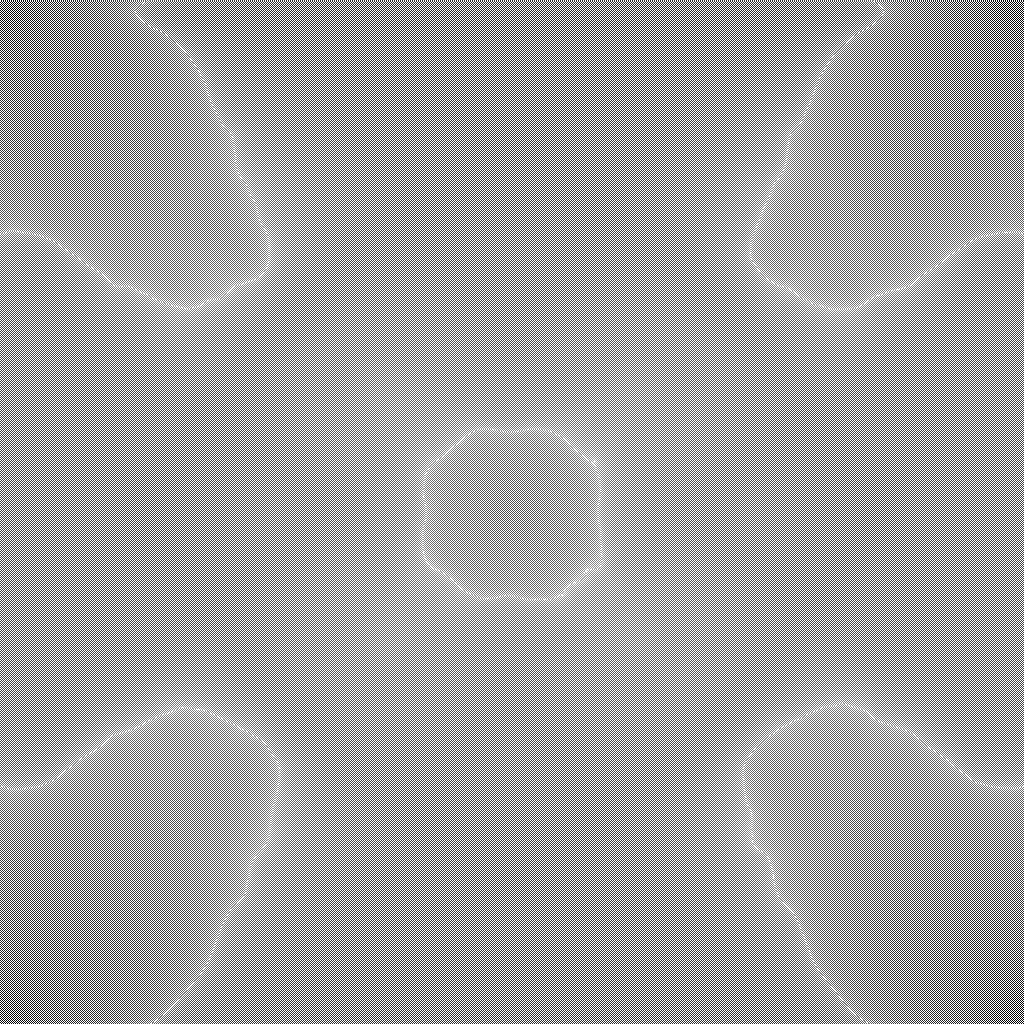}};

\node (shift) [application, right of=2Dfft, xshift=2.0cm] {\includegraphics[scale=0.22]{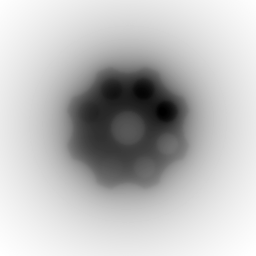}}; 
\draw [arrow] (Psino) -- node [above] {$0,1$} (zeroP);
\draw [arrow] (zeroP) -- node [left] {$2$} (bessel);
\draw [arrow] (bessel) -- node [above] {$3$} (1Dfft);
\draw [arrow] (1Dfft) --  node [above] {$4$} (kernel);
\draw [arrow] (kernel) -- node [left] {$5$} (interp);
\draw [arrow] (interp) -- node [above] {$6$} (2Dfft);
\draw [arrow] (2Dfft) -- node [above] {$7$} (shift);
\end{tikzpicture}
\caption{Theoretical fluxogram \textsc{bst} formula, from Eq.(\ref{eq:bst}).}
\label{fig:theoFlow}
\end{figure}

Each processing step $P_k$ can be implemented in a parallel form with \textsc{cuda}.
Step $P_0$ and $P_1$ indicates an interpolation to polar coordinates with the multiplication of
the Kaiser-Bessel window function, respectively. This is an easy process, computed with
complexity $O(N V)$. Step $P_2$ is the zero padding of
the polar sinogram - equivalent to an oversampling in the frequency domain - with
the same complexity of $P_0$. Even though $P_1$ and $P_2$ can be merged into one single
step, they were considered disjoint operations in our customized implementation. Step $P_3$ is a one-dimensional Fourier Transform of the data from step $P_2$. Step $P_4$ is the convolution of the polar sinogram with kernel $1/\sigma$. This part
was divided in $m$ parallel \textsc{fft}'s, each computed at an individual thread using advanced strategies with complexity $O(N \log N)$. Step $P_5$ is an interpolation from polar to cartesian coordinates in the frequency domain. The bigger the zero padding at step $P_2$, the better this part will behave, preventing aliasing artifacts. Step $P_6$ is a two-dimensional inverse Fourier transform of the data from step $P_5$. This is an operation with low computational complexity and obtained with order $O(N^2 \log N)$. Step $P_7$ is the \textsc{fft} shifting from previous step. We could add a $P_8$ step, an optional two-dimensional interpolation of the resulting image to the correct feature domain $[-1,1]\times[-1,1]$. A theoretical fluxogram for \textsc{bst}, describing processing steps $\{P_0, \ldots, P_7\}$ is presented in Figure \ref{fig:theoFlow}.


Computing the backprojection operator $\Back$ directly from equation (\ref{eq:back}) produce a high-computational complexity algorithm which is easy to implement in a parallel structure, either in \textsc{cpu} or \textsc{gpu}. Such approach has a computational complexity of $O(N^3)$ per slice, which in turns implies a total cost of $O(N^4)$ for a set of $N$ sinogram images $N$. We denote this as a \emph{Slant-stack} approach (\textsc{SS}) since summation is performed over straight lines with slope according to the input angle $\theta$ \cite{fslant}. A block of $Q$ sinograms is usually processed at once due to the fact that the backprojection kernel can be processed simultaneously for a tomographic scan using parallel rays. The fluxogram presented in Figure \ref{fig:theoFlow} indicates that each step of \textsc{bst} has computational complexity bounded by $O(N^2 \log N)$.

\section{Pipeline for Distribution}
\label{sec:pipeline}

\begin{figure}[h]
\centering
\includegraphics[scale=0.3]{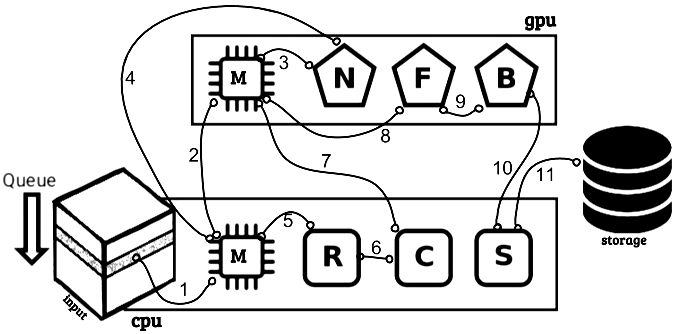}
\caption{Pipeline E(CPU/GPU) using a generic backprojection kernel {\bf B}.}
\label{fig:flowCpuGpu}
\end{figure}

The full implementation is composed of multiple different stages: sinogram normalization (N), low-pass filtering (F), backprojection (B), ring filtering (R), centering of sinograms (R) and saving reconstructed data to storage (S). A typical sequence for these operations is presented in Figure \ref{fig:flowCpuGpu}.
Although all stages operate on the same data, each has different characteristics regarding resource usage. Loading the input data and storing the output results are I/O bound operations, other stages are \textsc{cpu} bound and some are \textsc{gpu} bound. Even among the \textsc{gpu} bound stages, some are more compute intensive while others are more memory intensive.

Having such a heterogeneous set of process stages, it is desirable to optimize their execution in order to maximize resource utilization and consequently minimize execution time. Such optimization becomes non-trivial as the available resources increase and the scheduling possibilities are amplified. Although the stages must be executed in a given order for a given set of input data, they can be executed in parallel with different sets of input data, as long as each data set is processed by each stage in the expected order. To help with the scheduling of the available jobs (the execution of one of the stages on one of the input data sets), a helper framework was implemented. This framework consists of an abstract representation of a pipeline of interconnected stages. Data enters the pipeline and is processed by each stage sequentially until it reaches the end of the pipeline, producing the resulting output data for the given input data. Each stage is represented as a processing function that receives a generic data blob and returns a resulting generic data blob. The function therefore processes a data block into another data block. The contents of the data block is specific to each stage, and it is a requirement that the resulting output block is in the expected format for the input block of the stage that follows. With the implementation modeled as a pipeline, some resource optimizations can be applied generically. The first is the execution of each stage in a separate \textsc{cpu} thread. This is not an optimization to primarily use different \textsc{cpu} cores to execute stages in parallel, although it does contribute to the usage of more resources. Instead, the main advantage is that it allows the operating system to schedule the different stages depending on their resource usage patterns. This means that I/O bound stages can start, and when an I/O operation halts the execution, a stage that is \textsc{gpu} bound can be executed, and when that stage is waiting on the results from the \textsc{gpu} another stage can be executed or the idle stages can be resumed if the data they're waiting for is ready.

\begin{table}[h]
\tiny
\setlength\tabcolsep{1.0pt}
\centering
\begin{tabular}{cccccccc}
\hline
& \multicolumn{2}{c}{\bf Jetson TX1} & \multicolumn{2}{c}{\bf GT740M} & \multicolumn{2}{c}{\bf Titan X} & \\
\hline
\textsc{dataset} & \textsc{ss} & \textsc{bst} & \textsc{ss} & \textsc{bst} & \textsc{ss} & \textsc{bst} & \textsc{q} \\
\hline
$\sf 2048^3$ & & & & & 895.072083 & 190.321701 & 5 \\
& & & & & 799.828003 & 228.629852 & 10 \\
$\sf 1536^3$ & & & & & 174.356842 & 56.577358 & 5 \\
& & & & & 253.741974 & 71.588837 & 10 \\
$\sf 1024^3$ & & & 327.243317 & 62.341717 & 30.724861 & 17.413599 & 5 \\
& & & 259.171601 & 57.566849 & 24.224422 & 17.024681 & 10 \\
$\sf 512^3$ & 26.321959 & 13.171736 & 23.42721 & 9.229262 & 3.858362 & 4.774002 & 5 \\
& 24.451975 & 12.754734 & 18.632298 & 8.237438 & 3.802566 & 3.744505 & 10 \\
& 21.064373 & 13.432063 & 16.567259 & 7.892296 & 4.142621 & 3.781465 & 15 \\
$\sf 256^3$ & 3.516305 & 3.50024 & 1.948982 & 1.886878 & 1.022957 & 1.032447 & 5 \\
& 3.128555 & 3.290957 & 1.692543 & 1.437727 & 0.965807 & 0.969846 & 10 \\
& 3.041064 & 3.131222 & 1.567504 & 1.355728 & 0.965546 & 0.978986 & 15 \\
\hline
$\sf 2048^2 x 1024$ & & & & & 1986.611572 & 274.715823 & 5 \\
& & & & & 1172.36645 & 517.887451 & 10 \\
$\sf 1536^2 x 768$ & & & & & 72.134781 & 70.911224 & 5 \\
& & & & & 66.979645 & 28.303719 & 10 \\
$\sf 1024^2 x 512$ & & & 165.145828 & 42.321434 & 14.960188 & 9.268951 & 5 \\
& & & 130.446228 & 35.056644 & 14.851658 & 9.82362 & 10 \\
$\sf 512^2 x 256$ & 14.061594 & 8.30939 & 16.655327 & 5.380389 & 2.084591 & 2.029896 & 5 \\
& 13.922671 & 8.293776 & 9.346146 & 5.003119 & 1.998881 & 2.203163 & 10 \\
& 16.117891 & 8.202461 & 8.63123 & 4.846304 & 1.995749 & 1.907305 & 15 \\
$\sf 256^2 x 128$ & 2.515128 & 2.740289 & 1.0246685 & 1.205075 & 0.643044 & 0.681087 & 5 \\
& 2.421637 & 2.650361 & 1.036608 & 1.056225 & 0.507826 & 0.525796 & 10 \\
& 2.237847 & 2.310273 & 0.957721 & 1.028077 & 0.469525 & 0.500501 & 15 \\
\hline
\end{tabular}
\caption{Execution time (in seconds) for the pipeline E(GPU), using only one 
\textsc{gpu}, and comparing backprojection kernels SS and BST.}
\label{tab:E1}
\end{table}

With each stage executing in different threads, the jobs are executed concurrently. An optimization opportunity arises in managing the number of jobs in flight. In one extremity, the number of jobs in flight can be limited to one, which effectively makes the pipeline run sequentially with no concurrency. On the other extremity, the number can be unlimited. While this might appear to be the optimal solution, because generally stages don't have the same execution time, some stages end up being bottlenecks. When this occurs, many jobs remain queued before the slowest stage, and they can potentially be holding resources that they are not using while idle. An example of this was that between two \textsc{gpu} specific stages, the data transferred between them resided in the \textsc{gpu} memory. Because the second stage was slower than the first stage, the \textsc{gpu} ran out of memory due to the amount of queued jobs for the second stage. The solution to this problem was to allow limiting the number of jobs in queue before each stage. Each stage can have its maximum queue length individually tuned, to allow limiting only the jobs in flight that may strain resources while idle.

\begin{figure}[h]
\centering
\includegraphics[scale=0.3]{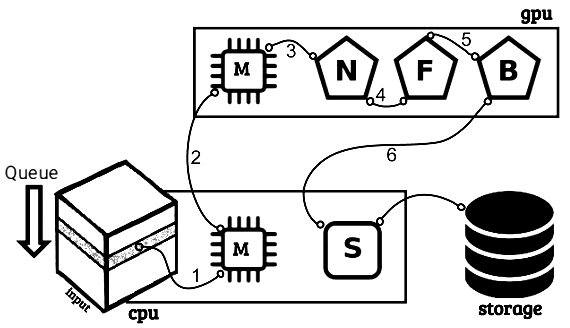}
\caption{Pipeline for \textsc{gpu} code using a generic backprojection kernel {\bf B}.}
\label{fig:flowGpu}
\end{figure}

The final generic optimization is to allow running the same stage in a pool of threads. If the execution isn't fully using a specific resource, the stage that uses it most can be executed in more than one thread, which means they use that resource concurrently, potentially increasing its utilisation. An example is \textsc{cpu} core usage. If not all of the systems core are in use, have the same stage execute in more than one thread allows the operating system to schedule more threads on the \textsc{cpu}, attempting to maximize its usage. This also makes it easier when using multiple \textsc{gpu}s, since each thread can be assigned to work with a single \textsc{gpu}. Even having more than one thread per \textsc{gpu} can increase throughput, because even if the task is the same, the \textsc{gpu} driver can schedule data transfers between the main memory and the \textsc{gpu} memory while the \textsc{gpu} is actively running another job (this is allowed in \textsc{cuda} as long as the program is configured to have a per-thread default stream).

Such abstract implementation of a work pipeline allows the algorithm to be broken up into stages - as shown in Figure \ref{fig:flowCpuGpu} - and makes it easier to manage the execution of these stages in order to maximize resource usage. After adapting the implementation to the pipeline, some execution parameters can be tweaked to increase throughput. These include the number of threads for each stage, and the maximum queue size before each stage. This allows a simple way to experiment with different implementations by adapting the work done in each stage, and also quick experimentation with job scheduling to validate how different allocations result in different resource usage and performance.

\begin{table}[h]
\centering
\footnotesize
\setlength\tabcolsep{2.0pt}
\begin{tabular}{c|c|c|c|c|c|c|c|c}
\hline
\multicolumn{2}{c|}{\textsc{jetson tx1}} & \multicolumn{3}{c|}{\textsc{gt}740\textsc{m}} & \multicolumn{4}{c}{\textsc{titan x}} \\
\hline
$256^3$ & $512^3$ & $256^3$ & $512^3$ & $1024^3$ & $256^3$ & $512^3$ & $1024^3$ & $2048^3$ \\
\hline
1.6 & 6.80 & 0.56 & 2.34 & 10.52 & 0.69 & 3.02 & 14.02 & 110.81 \\
\hline
\end{tabular}
\caption{Average time (in seconds) expended reading the input data $\bm Y$ using a non-optimized HDF call. All executions were done using blocksize of $Q=10$.}
\label{tab:read1}
\end{table}

\newcommand{\mysize}{0.25}
\begin{figure}[t]
\centering
\begin{tabular}{cc}
(a.1) & (a.2) \\
\includegraphics[scale=\mysize]{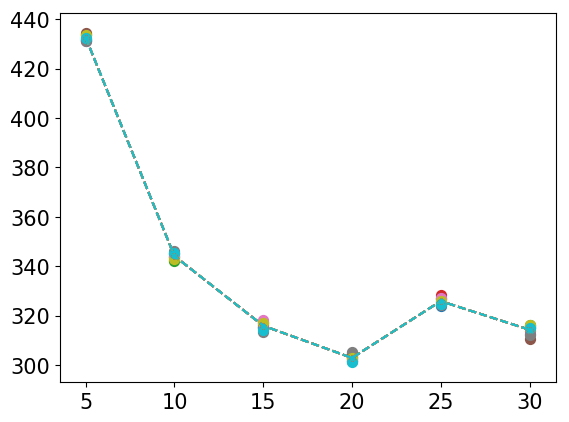} & \includegraphics[scale=\mysize]{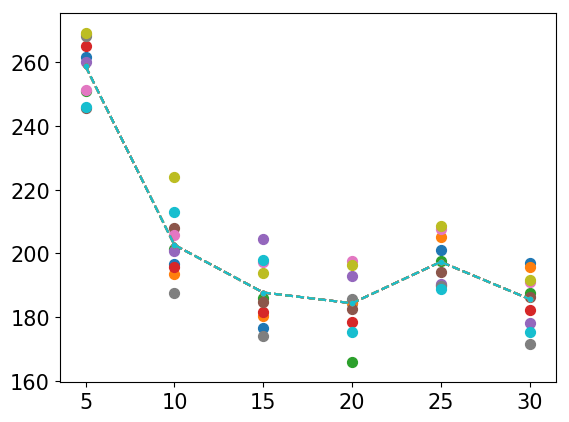} \\
(a.3) & (a.4) \\
\includegraphics[scale=\mysize]{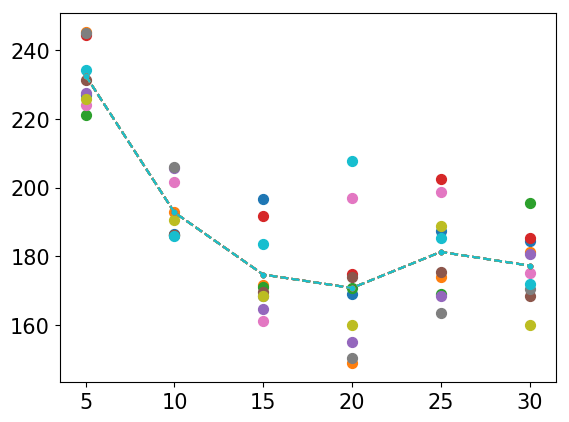} & \includegraphics[scale=\mysize]{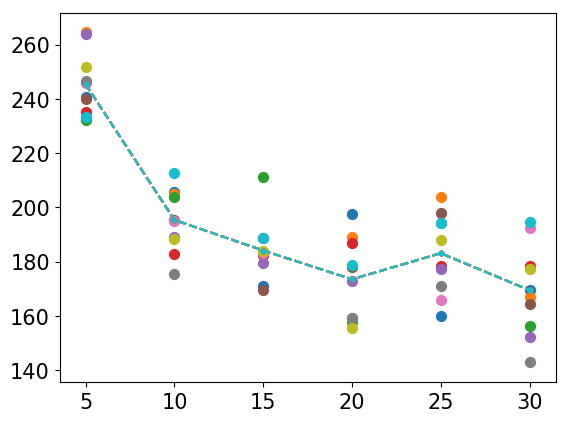} \\
(b.1) & (b.2) \\
\includegraphics[scale=\mysize]{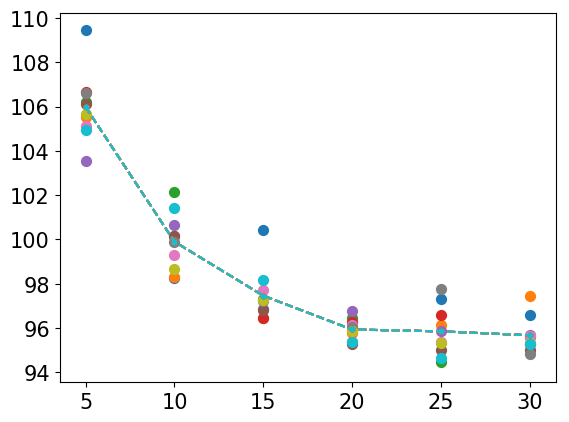} & \includegraphics[scale=\mysize]{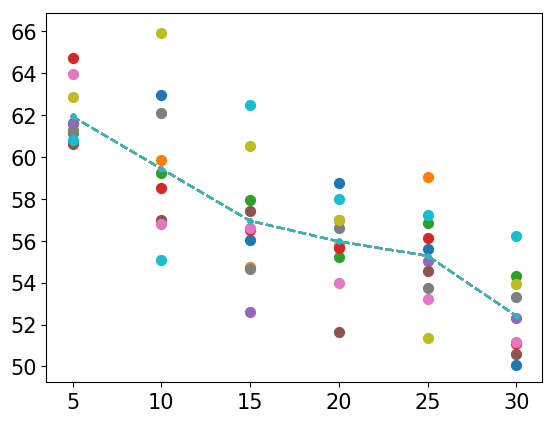} \\
(b.3) & (b.4) \\
\includegraphics[scale=\mysize]{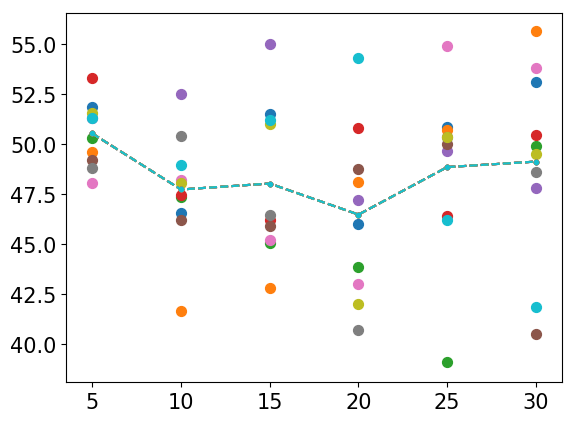} & \includegraphics[scale=\mysize]{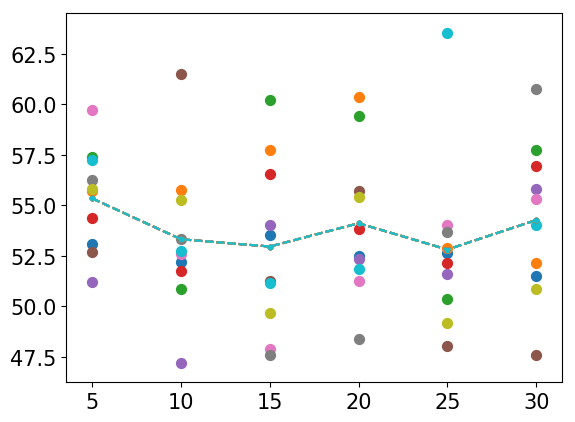}
\end{tabular}
\caption{\footnotesize Execution time (in seconds) for pipeline E(GPU) as a function of blocksize $Q$, using 1,2,3,4 GPUs and the high-complexity kernel SS. (a) SGI/C2108-GP5 (b) Minsky/IBM.}
\label{fig:SSgpu}
\end{figure}

\begin{figure}[t]
\centering
\begin{tabular}{cc}
(a.1) & (a.2) \\
\includegraphics[scale=\mysize]{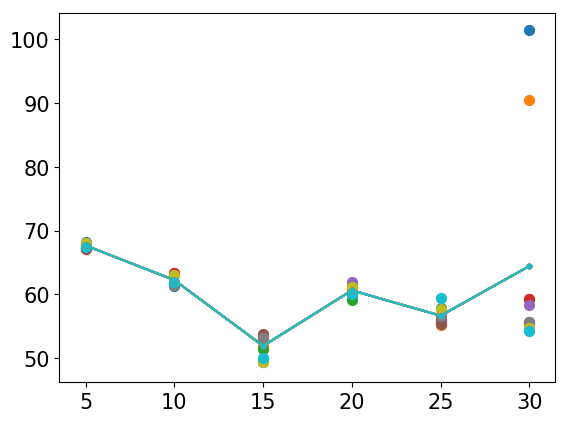} & 
\includegraphics[scale=\mysize]{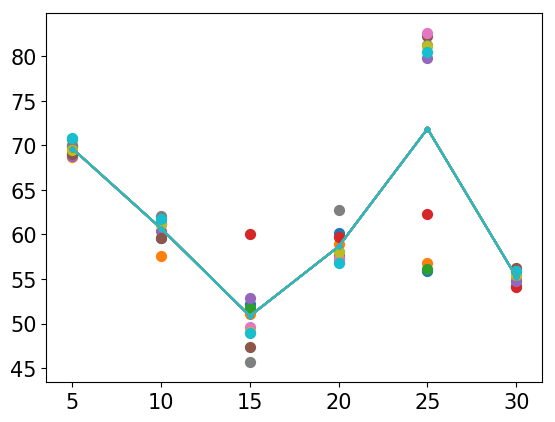} \\
(a.3) & (a.4) \\
\includegraphics[scale=\mysize]{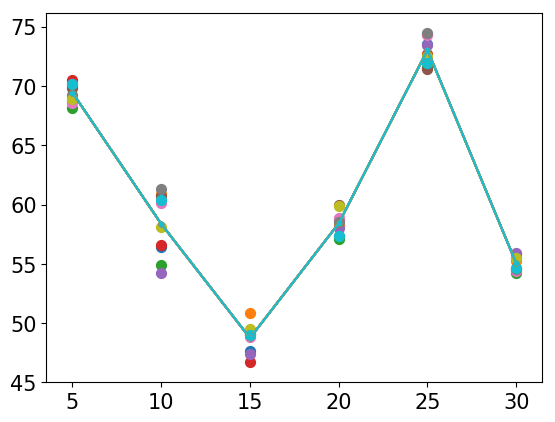} & 
\includegraphics[scale=\mysize]{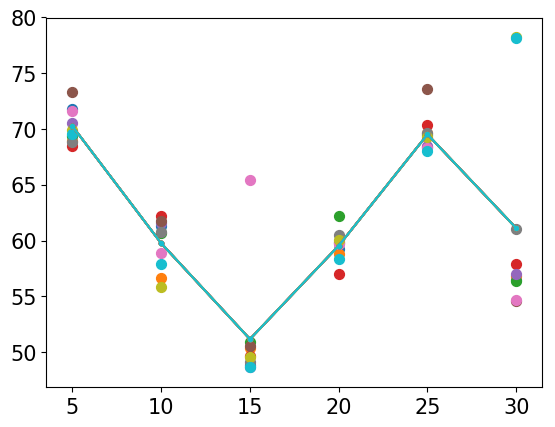} \\
(b.1) & (b.2) \\
\includegraphics[scale=\mysize]{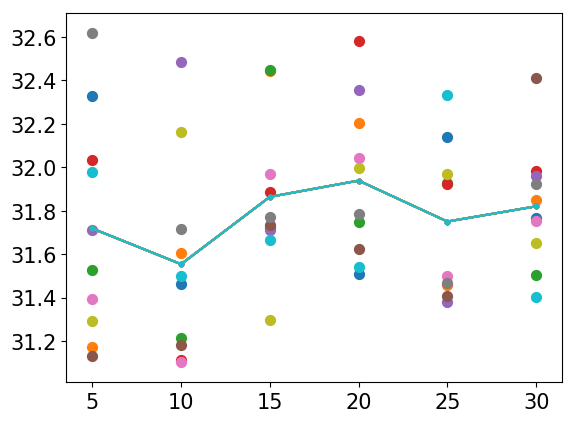} & 
\includegraphics[scale=\mysize]{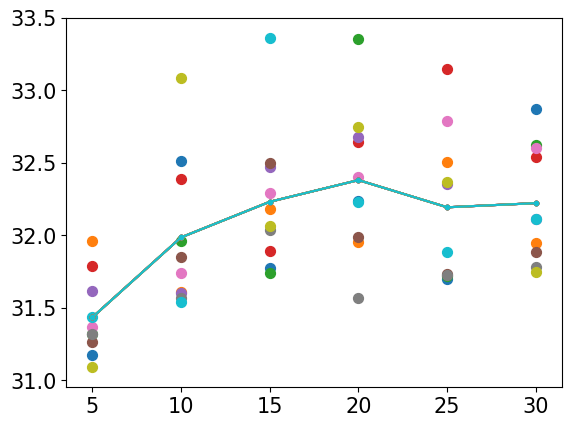} \\
(b.3) & (b.4) \\
\includegraphics[scale=\mysize]{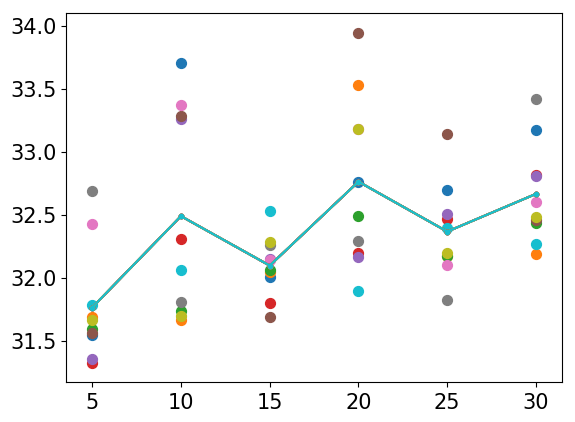} & 
\includegraphics[scale=\mysize]{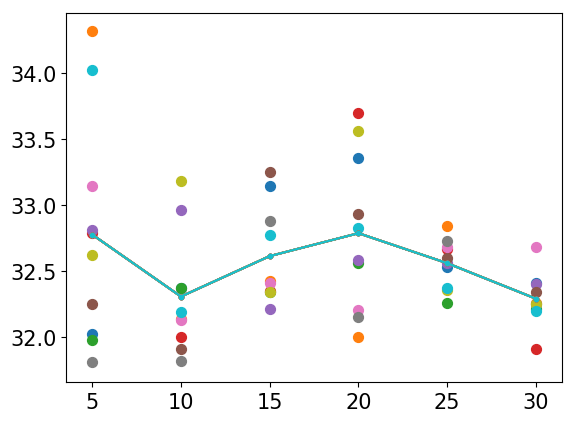}
\end{tabular}
\caption{\footnotesize Execution time (in seconds) for pipeline E(GPU) as a function of blocksize $Q$, using 1,2,3,4 GPUs and the low-complexity kernel BST. (a) SGI/C2108-GP5 (b) Minsky/IBM}
\label{fig:BSTgpu}
\end{figure}

\section{Performance}

\begin{figure}[t]
\centering
\begin{tabular}{cc}
(a) & (b) \\
\includegraphics[scale=\mysize]{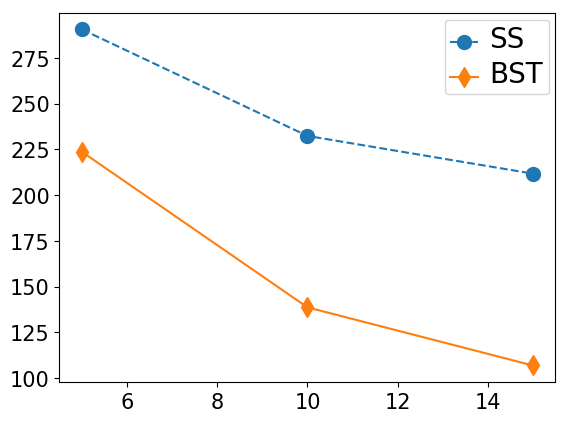} & 
\includegraphics[scale=\mysize]{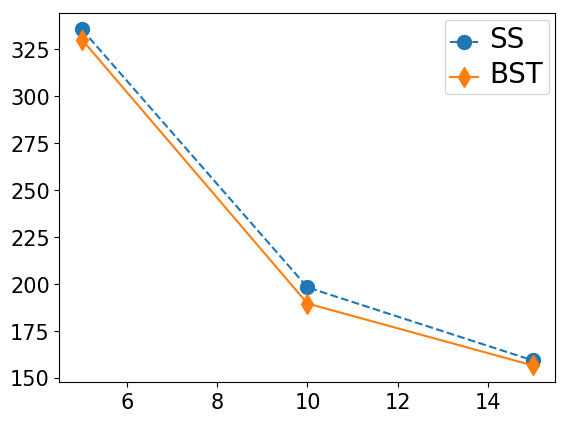} \\
\end{tabular}
\caption{\footnotesize Execution time (in seconds) for pipeline E(CPU/GPU) as a function of blocksize $Q$, using 4 GPUs (a) SGI/C2108-GP5 (b) Minsky/IBM.}
\label{fig:full}
\end{figure}

Our pipeline, as presented in Section \ref{sec:pipeline}, is dependent on the time for reading the input data $\bm Y$ - see definition on Figure \ref{fig:dataset}.  The hierarchical data format \textsc{hdf5} is becoming very popular for the storage of measured data at synchrotron facilities. In this sense, our algorithm is evaluated in two ways: {\bf (a)} Extracting - without \textsc{mpi} capabilities - a block of $Q$ images. Here, we have used the standard call of \textsc{hdf} to read the given number of images. The block of images is then processed following the pipeline of Figures \ref{fig:flowCpuGpu} and \ref{fig:flowGpu}, except that we are not concerned with output. Hence, the flux of reconstructed data to storage is not considered in our analysis. Executions using the pipeline of Figure \ref{fig:flowCpuGpu} are refered as \textsc{e(cpu/gpu)} as they have an extra processing of parallel kernels {\bf R} and {\bf C} at \textsc{cpu}. Executions using the pipeline of Figure \ref{fig:flowGpu} are refereed as \textsc{e(gpu)}. The pipeline was executed using $W=8$ work items, each using $T = 2 \times G$ threads, being $G$ the number of availables \textsc{gpu}s. 
Needless to say that $Q$ is bounded by the \textsc{gpu} memory and therefore by $G$ and $W$, i.e., $A_{\mbox{\footnotesize\bf N}}(Q W T) + A_{\mbox{\footnotesize\bf F}}(Q W T)  + A_{\footnotesize \mbox{\bf B}}(QWT) \leq \mbox{\bf M}$ with $A_{\mbox{\bf N}}$ the amount of memory allocation for kernel {\bf N} and {\bf M} the global memory of a single \textsc{gpu}. {\bf (b)} Measuring the elapsed time of the pipeline just reading the input data $\bf Y$, and turning off all subsequent kernels. In average, we understand that the execution time of the reconstruction is bounded by this step of the pipeline. We denote this execution as \textsc{r(cpu/gpu)}.

Table \ref{tab:E1} presents the elapsed times for the pipeline \textsc{e(gpu)} using only one \textsc{gpu} and standard tomographic dimensions\footnote{We have adopted a three-dimensional dataset described by an ellipsoid. The feature image is defined as $\bm x(u) = \rho$ if $(u_1/A)^2 + (u_2/B)^2 \leq 1 - (s/C)^2$ and zero otherwise. Here, $A$ and $B$ are parameters defining the shape of the semi-major axis $u_1,u_2$ of the ellipse and $C$ defining the length of slice axis $|s| \leq C$. The Radon transform of $\bb x$ can be obtained analytically \cite{kak_slaney}. A discretization of the sinogram using $\{N,V\}$ points for $t$ and $\theta$ axis respectively, gives us a three-dimensional dataset $\bm Y$.}. Using \textsc{bst} with a powerful \textsc{gpu} like \textsc{titan x} we can achieve a feasible reconstruction of a dataset of dimension $2048^3$ within 3 minutes. At the same dimension
and after 10 executions of the pipeline \textsc{r(cpu/gpu)} with a blocksize of $Q=10$, we 
obtain an average of $107$ seconds wasted on reading the input data - see Table \ref{tab:read1}. Hence, the reconstruction can be achieved within one minute. Using the same reasoning for the other devices, 
the reconstruction with \textsc{jetson} runs within 6 seconds for a dimension of $512^2$ and 
within 1 second for dimensions $512^2 \times 256$. The standard \textsc{gpu-gt740} runs a reconstruction within 46 seconds for a dimension $1024^3$. 

\begin{table}[h]
\centering
\tiny
\setlength\tabcolsep{1.0pt}
\begin{tabular}{|c|c|c|c|c|c|c|c|c|c|c|c|c|}
\hline
& \multicolumn{6}{c|}{\textsc{minsky}/\textsc{ibm}} & \multicolumn{6}{c|}{\textsc{sgi/c2108-gp5}} \\
\hline
Q & 5 & 10 & 15 & 20 & 25 & 30 & 5 & 10 & 15 & 20 & 25 & 30 \\
\hline
1\textsc{gpu} & 30.3 & 30.10 & 30.46 & 30.38 & 29.92 & 29.85 & 38.70 & 41.81 & 38.45 & 42.47 & 39.41 & 41.77 \\
\hline
2\textsc{gpu}s & 30.37 & 29.97 & 30.29 & 30.16 & 30.29 & 30.21 & 39.13 & 40.89 & 40.96 & 38.95 & 39.21 & 41.87 \\
\hline
3\textsc{gpu}s & 30.70 & 30.37 & 30.51 & 30.37 & 30.60 & 30.30 & 35.92 & 37.73 & 40.54 & 39.66 & 41.94 & 42.45 \\
\hline
4\textsc{gpu}s & 30.58 & 30.75 & 30.55 & 30.67 & 30.20 & 30.61 & 35.99 & 37.52 & 42.43 & 39.68 & 42.88 & 42.15 \\
\hline
\end{tabular}
\caption{Average time (in seconds) expended reading the input data $\bm Y$ using a non-optimized HDF call, different blocksizes and two clusters.}
\label{tab:read}
\end{table}

A large number of executions (at two different clusters) of the pipeline \textsc{e(gpu)} using real data with dimensions $(N,V)=(2048,2000)$  is presented in Figures \ref{fig:SSgpu} and \ref{fig:BSTgpu}. Plots (a.$k$) therein represents 10 elapsed times (in seconds) versus the blocksize number $Q$ - with $k$ \textsc{gpu}s for code distribution running at a SGI/C2108-GP5 server coupled with 4 \textsc{nvidia} K80. Plots (b.$k$) represents 10 elapsed times (in seconds) versus the blocksize number $Q$ - with $k$ \textsc{gpu}s for code distribution running at a Minsky/IBM server coupled with 4 \textsc{nvidia} P100. Times for \textsc{bst} (Fig.\ref{fig:BSTgpu}) are much lesser than \textsc{ss} (Fig.\ref{fig:SSgpu}), as predicted by theory. The average time expended reading the input data for these servers using a non-optimized \textsc{hdf} call, is presented in Table \ref{tab:read} for different values of  blocksize $Q$.  The processing time expected for \textsc{bst} is near 1 second using 1,2,3 or 4 \textsc{gpus} at Minsky cluster while 30 seconds for the \textsc{sgi} cluster. The average time for a complete reconstruction using the high-complexity kernel \textsc{ss} depends on the number of \textsc{gpu}s used for distribution. The elapsed times for the hybrid pipeline \textsc{e(cpu/gpu)} - see Figure \ref{fig:flowCpuGpu} - running on the same data is presented in Figure \ref{fig:full}. Now, it is clear that the time expended processing the data on the \textsc{cpu} affect performance dramatically. In fact, the \textsc{cpu} kernel {\bf R} process a single slice using a Conjugate gradient method for stripe suppression \cite{miquelesRings}. Processing a block of $Q$ images, although easy to implement, does not present the same quality of sinogram restoration because correction is slice dependent. Hence, even with a $O(N)$ computational complexity, the ring suppression plus data transfer from \textsc{cpu/gpu}
should be avoided in this pipeline. There is certainly a huge space for improvements on the
ring suppression algorithm, handling a block of sinograms. The \textsc{cpu} kernel {\bf C} - for centering sinograms - is optional and easy to implement for a block of images not affecting the final performance. 



\section{Conclusions}

Fast reconstructions of large tomographic datasets is feasible using the \textsc{bst}
formula (\ref{eq:bst}). The superscalar pipeline - running only \textsc{gpu} kernels with computational complexity bounded above by $O(N^2 \log N)$ - is able to complete a
\textsc{3d} reconstruction within 1 second of running time provided the input dataset is transferred completely to the memory. New imaging detectors capable to store the data
in a local buffer could benefit from this reconstruction approach. We emphasize that 
other low-complexity algorithms \cite{andersson,gursoy2014tomopy,marone2017towards}
are also capable to provide fast results. Nonetheless, a fast backprojector (\textsc{bst}) and a fast projector (\textsc{fst}) implemented in the same pipeline presented here, give us the chance to implement advanced reconstruction strategies \cite{helou2017superiorization} (mostly iterative) with the same low computational complexity. Further improvements on the code\footnote{Download: \texttt{https://github.com/exmiqueles/raft.git}} are under progress which include: usage of \textsc{mpi} strategies to read the data, ring-filtering kernel to process a block of sinograms, thread optimization among others. 

\medskip

\subsection*{Acknowledgments}

\medskip
We would like to thank the \textsc{ibm} team for providing access to the 
Poughkeepsie Benchmarking Technical Computing Cloud: Douglas M. Dreyer, Victoria Nwobodo, James Kuchler, Khajistha Fattu, Antonio C.Navarro and Leonardo A.G. Garcia. Thanks also to Harry Westfahl Jr. for many valuable suggestions. The \textsc{titan x} used for this research was donated by the NVIDIA Corporation.
{
\footnotesize
\bibliographystyle{unsrt}
\bibliography{main}

\begin{thebibliography}{10}

\bibitem{kak_slaney}
Avinash~C Kak and Malcolm Slaney.
\newblock {\em Principles of computerized tomographic imaging}.
\newblock IEEE press, 1988.

\bibitem{natterer}
Frank Natterer.
\newblock {\em The mathematics of computerized tomography}, volume~32.
\newblock Siam, 1986.

\bibitem{deans}
Stanley~R Deans.
\newblock {\em The Radon transform and some of its applications}.
\newblock Courier Corporation, 2007.

\bibitem{andersson}
Fredrik Andersson.
\newblock Fast inversion of the radon transform using log-polar coordinates and
  partial back-projections.
\newblock {\em SIAM Journal on Applied Mathematics}, 65(3):818--837, 2005.

\bibitem{bresler2}
Ashvin George and Yoram Bresler.
\newblock Fast tomographic reconstruction via rotation-based hierarchical
  backprojection.
\newblock {\em SIAM Journal on Applied Mathematics}, 68(2):574--597, 2007.

\bibitem{miqueles}
Eduardo Miqueles and Elias~S Helou.
\newblock Fast backprojection operator for synchrotron tomographic data.
\newblock In {\em European Conference on Mathematics for Industry}. Springer,
  2014.

\bibitem{marone}
F~Marone and M~Stampanoni.
\newblock Regridding reconstruction algorithm for real-time tomographic
  imaging.
\newblock {\em Journal of synchrotron radiation}, 19(6):1029--1037, 2012.

\bibitem{gursoy2014tomopy}
Doga G{\"u}rsoy, Francesco De~Carlo, Xianghui Xiao, and Chris Jacobsen.
\newblock Tomopy: a framework for the analysis of synchrotron tomographic data.
\newblock {\em Journal of synchrotron radiation}, 21(5):1188--1193, 2014.

\bibitem{marone2017towards}
Federica Marone, Alain Studer, Heiner Billich, Leonardo Sala, and Marco
  Stampanoni.
\newblock Towards on-the-fly data post-processing for real-time tomographic
  imaging at tomcat.
\newblock {\em Advanced Structural and Chemical Imaging}, 3(1):1, 2017.

\bibitem{shkarin2015gpu}
Roman Shkarin, Evelina Ametova, Suren Chilingaryan, Timo Dritschler, Andreas
  Kopmann, Alessandro Mirone, Andrei Shkarin, Matthias Vogelgesang, and Sergey
  Tsapko.
\newblock Gpu-optimized direct fourier method for on-line tomography.
\newblock {\em Fundamenta Informaticae}, 141(2-3):245--258, 2015.

\bibitem{mirone2010pyhst}
A~Mirone, R~Wilcke, A~Hammersley, and C~Ferrero.
\newblock Pyhst--high speed tomographic reconstruction, 2010.

\bibitem{miquelesRings}
Eduardo~X Miqueles, Jean Rinkel, Frank O'Dowd, and JSV Berm{\'u}dez.
\newblock Generalized titarenko's algorithm for ring artefacts reduction.
\newblock {\em Journal of synchrotron radiation}, 21(6):1333--1346, 2014.

\bibitem{potts}
Daniel Potts and Gabriele Steidl.
\newblock New fourier reconstruction algorithms for computerized tomography.
\newblock In {\em International Symposium on Optical Science and Technology},
  pages 13--23. International Society for Optics and Photonics, 2000.

\bibitem{fslant}
A~Averbuch, RR~Coifman, DL~Donoho, M~Israeli, and J~Walden.
\newblock {\em Fast Slant Stack: A notion of Radon transform for data in a
  cartesian grid which is rapidly computible, algebraically exact,
  geometrically faithful and invertible}.
\newblock Department of Statistics, Stanford University, 2001.

\bibitem{helou2017superiorization}
Elias~S Helou, Marcelo~VW Zibetti, and Eduardo~X Miqueles.
\newblock Superiorization of incremental optimization algorithms for
  statistical tomographic image reconstruction.
\newblock {\em Inverse Problems}, 33(4):044010, 2017.

\end{thebibliography}
}


\end{document}